%% file: main.tex
\documentclass{article}
\usepackage{float}
\usepackage[utf8]{inputenc}
\usepackage[caption = false]{subfig}
\usepackage{spconf,amsmath,graphicx}
\usepackage{xcolor}
\usepackage{hyperref}
\usepackage{lipsum,multicol}

\usepackage{amsfonts}
\usepackage{ulem}
\usepackage{tikz}
\usepackage{stmaryrd}
\usepackage{amsthm}
\usepackage{enumitem}
\usetikzlibrary{automata, arrows.meta, positioning}
\newtheorem{proposition}{Proposition}
\usepackage{enumitem}
\graphicspath{ {img/} }

\title{MQ-coder inspired arithmetic coder for synthetic DNA data storage}
\name{Xavier Pic$^{\star}$, Melpomeni Dimopoulou$^{\star}$, Eva Gil San Antonio$^{\star}$, Marc Antonini$^{\star}$}
\address{$^{\star}$ I3S laboratory, Côte d’Azur University and CNRS\\UMR 7271, Sophia Antipolis, France}
\begin{document}
%
\maketitle
\input{abstract}
\vspace{-1\baselineskip}
\input{intro}
\vspace{-1\baselineskip}
\input{context}

\vspace{-1\baselineskip}
\input{arithmetic_coder}
\vspace{-1\baselineskip}
\input{results}
\vspace{-1\baselineskip}
\input{conclusion}
\newpage
\bibliographystyle{IEEEbib}
\bibliography{strings}

\end{document}

%% file: abstract.tex
\begin{abstract}
\par

Over the past years, the ever-growing trend on data storage demand, more specifically for "cold" data (i.e. rarely accessed), has motivated research for alternative systems of data storage.
Because of its biochemical characteristics, synthetic DNA molecules are now considered as serious candidates for this new kind of storage. This paper introduces a novel arithmetic coder for DNA data storage, and presents some results on a lossy JPEG 2000 based image compression method adapted for DNA data storage that uses this novel coder. 

The DNA coding algorithms  presented here have been designed to efficiently compress images, 
encode them into a quaternary code, and finally store them into synthetic DNA molecules. 
This work also aims at making the compression models better fit the problematic that we encounter when storing data into DNA, 
namely the fact that the DNA writing, storing and reading methods are error prone processes. 

The main take away of this work is our arithmetic coder and it's integration into a performant image codec.

\end{abstract}

%% file: intro.tex
\section{Introduction}
\vspace{-2mm}
\label{sec:intro}
\par

Images represent a large part of the cold data stored in data centers. For that reason, it is crucial to develop image coders that have a good compression performance and are adapted to the problematics inherent to DNA data storage. Over the past years, some coders \cite{Church, Goldman2013, Aeon}, were developed to encode data into DNA. A research group has even been constituted to provide image compression solutions for DNA data storage: the JPEG DNA Ad Hoc group\footnote{\color{blue}\url{https://jpeg.org/jpegdna/index.html}}. Some research has aimed at having a good noise robustness, other works put their focus on the development of standards adapted to the constraints of image coding adapted to synthetic DNA data storage. In this paper, we present a new arithmetic coder for DNA data storage. We then adapt with it a performant image compression algorithm, JPEG 2000 \cite{JPEG2000} to make it work for DNA data storage. This paper first describes some concepts around DNA data storage and the JPEG 2000 codec, then explains the necessity of an arithmetic coder for DNA data storage, especially for image coding. Then it describes our proposed arithmetic coder and finally shows results on the integration of this new coder in an image codec.

%% file: context.tex
\vspace{-2mm}
\section{Context}
\vspace{-2mm}
\subsection{DNA data storage}
\vspace{-2mm}
\par 
Today, the digital world relies on increasingly large amounts of data, sotred over periods of time ranging from a few years to several centuries. Current tools are no longer sufficient and it is necessary to look for new storage solutions. One of the most promising solutions is to store information in the form of DNA, molecules that provide a very dense ---petabytes of data can be stored in a single gram of DNA--- and stable ---over very long periods when stored in the right conditions--- storage medium \cite{DNARAM}.
\par

The process of DNA data storage starts by encoding data into a quaternary stream composed of the 4 DNA symbols or nucleotides (A, C, T, and G). The encoded data is then physically synthesised into DNA molecules that are then stored into a safe environment. When data has to be read back, the stored molecules are amplified so to obtain many copies of each original sequence, or oligo, and the content is decyphered using DNA sequencers. This data is then decoded back into the binary file usually manipulated. However, the biochemical processes involved introduce some coding constraints \cite{Church} that, if not respected, dramatically increase the probability of an error (insertion, deletion and/or substitution) occurring. These constraints comprise avoiding homopolymers, patterns and unbalanced GC content.

Although constrained coding helps reducing the apparition of errors, it does not ensure a complete error free coding. To tackle the problems of errors, an important number of researchers \cite{Welzel, Imperial, DNASmart} are currently studying the development of accurate error models for synthetic DNA data storage.
\par
Algorithms specifically designed for DNA data storage generally show better reliability \cite{Goldman2013}. Because of the high cost of DNA synthesis, it is also important to take advantage of optimal compression, which can be achieved before synthesizing the sequence into DNA. As an example, among other relevant works, in \cite{DNAcoding} authors proposed a Discrete Wavelet Transform (DWT) image decomposition where the DWT coefficients are scalar quantized and then encoded using a quaternary code.
Over the past years, some methods have been developped to adapt to different sources \cite{JPEGDNABCT}, to improve the compression rate of this algorithm \cite{SFC4} and robustify it agianst noise \cite{MWCC2022Iulia, MWCC2022Xavier}.

This paper proposes a novel solution based on the JPEG 2000 image codec to encode images into DNA, More specifically, we will replace the MQ-coder with a DNA-like quaternary stream instead of the usual compressed binary stream that JPEG 2000 provides.
\par
\vspace{-1mm}
\subsection{The JPEG 2000 image compression algorithm}
\begin{figure}
    \centering
    \includegraphics[scale=0.45]{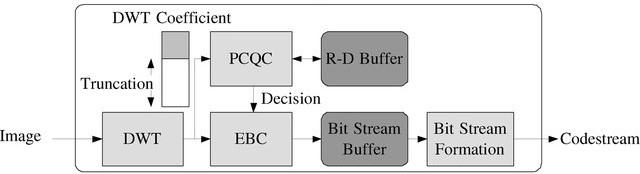}
    \vspace{-2mm}
    \caption{\footnotesize{Encoding schema of the DWT coefficients and its rate control by truncation of the coded DWT coefficients}}
    \label{fig:encodgin_schema}
\end{figure}

\vspace{-1mm}
\label{sec:JPEG2000}
JPEG 2000 is an image codec based on the encoding of Discrete Wavelet Transform coefficients \cite{DWT}. When it was first developped, it aimed at improving the performance of the original JPEG algorithm. The codec's algorithm is comprises different tasks: tiling, transform, quantization and encoding.
\begin{figure}
    \centering
    \includegraphics[scale=0.35]{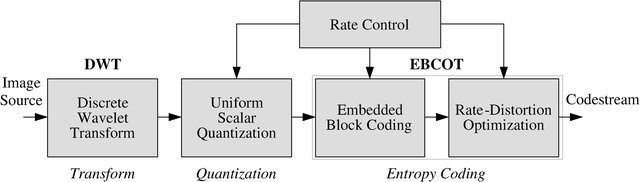}
    \vspace{-2mm}
    \caption{\footnotesize{The JPEG 2000 codec different steps: transofrm, quantization, bit plane representation and arithemtic coding}}
    \label{fig:JP2codec}
\end{figure}
The tiling simply cuts the image into seperate tiles that will be encoded separately.
The DWT transforms the signal in each tile into a set of coefficients, which are then scalar quantized. Although quantization has a negative impact on the reconstruction quality, it allows us to increase the compression performance. Moreover, the quantization step can be adjusted to improve quality at the expense of compression or vice-versa.
The encoding step (Fig. \ref{fig:encodgin_schema}) of the quantized coefficients is operated thanks to a series of algorithms, namely the bit-plane coder and the MQ-coder \cite{EBCOT}. The bit-plane coder first separates the coefficients by bit-planes, each bit-plane describing the bit of a certain weight. The MQ-coder then encodes these bit planes.

%% file: arithmetic_coder.tex
\vspace{-2mm}
\section{Motivations}
\vspace{-2mm}
\subsection{Image compression for DNA data storage}
\begin{figure*}[!ht]
    \centering
    \includegraphics[scale=0.185]{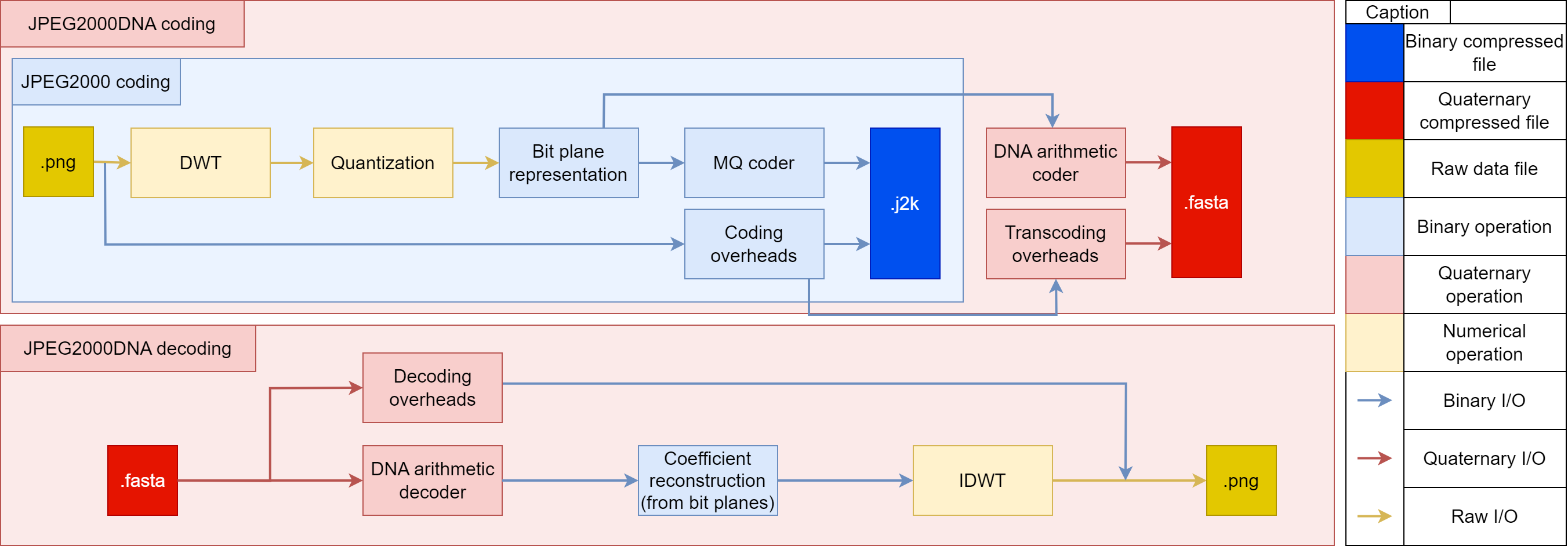}
    \caption{\footnotesize{General concept of the encoding and decoding scheme of the JPEG2000 compression algorithm adapted to DNA}}
    \label{fig:Codec}
\vspace{-0.5\baselineskip}
\end{figure*}

\vspace{-2mm}
The general idea of our proposed encoding process is depicted in Fig. \ref{fig:Codec} and can be very roughly described by the following steps. The input image has to be transformed into a set of coefficients using a DWT. The goal of this codec is to efficiently encode these coefficients rather than encoding the raw data. The encoding will be done thanks to a combination of coders based on the JPEG 2000 codec. The JPEG 2000 codec uses an algorithm described in Fig. \ref{fig:JP2codec} and in section \ref{sec:JPEG2000}. In our case, the JPEG 2000 codec has been modified to retrieve a bit-plane representation (result of the EBCOT coder) of each DWT coefficient that we will encode into a quaternary stream. In the end, this bit-plane representation is encoded using a constrained quaternary arithmetic coder using the $\{A,C,T,G\}$ alphabet. Decoding is ensured thanks to a decoder that has also been adapted to this constrained coding.\par
\vspace{-2mm}
\subsection{Arithmetic coder for DNA-like data}
\vspace{-2mm}
Arithmetic coders are a widely used method for coding data streams. They offer good performance for source codes that are non i.i.d.. Adapting these kinds of coders to DNA data storage would be an interesting step forward for the whole domain, and for the capabilities and applications it offers.
Moreover, arithmetic coders are used in specific image codecs such as JPEG 2000, that are famous for their good compression performance. The motivation for this new DNA coder was to develop an arithmetic coder adapted to the constraints of DNA data storage and integrate it into a JPEG 2000-based image codec for synthetic DNA data storage.
\vspace{-2mm}
\section{Principle of arithmetic coding}
\vspace{-2mm}
Generally, an arithmetic coder is used to re-encode a stream of symbols ---usually bits--- into a more compressed stream of symbols, that are also usually bits. The coder first associates the sequence of input symbols to a unique interval in $[0,1)$ that represents the input sequence. The more probable the sequence, the larger the interval it is represented by. \\
For a binary source, like the bit-plane representaiton of the DWT coefficients, the algorithm's main steps are:
\begin{enumerate}[itemsep=1pt,parsep=1pt]
    \item The interval $[x_{min}, x_{max})$ is initialised as $[0,1)$.
    \item For every received bit, two steps are performed:
        \begin{enumerate}[itemsep=1pt,parsep=1pt,label=2.\arabic*]
            \item The interval $[x_{min}, x_{max})$ is subdivided into two intervals. 
            \begin{itemize}[itemsep=1pt,parsep=1pt]
                \item The left interval is related to a received bit of value 0, and the right one to a received bit of value 1.
                \item The size of the two intervals is proportional to the probability of receiving this value for this bit (0 or 1).
            \end{itemize}
            \item The interval corresponding to the value of the received bit is selected, and replaces the current interval.
        \end{enumerate}
    \item The previous process is repeated for a fixed amount of bits received. 
    \item A mechanism to specify the end of the coded stream has to be put in place.
\end{enumerate}

\begin{figure}
    \centering
    \includegraphics[width=9cm]{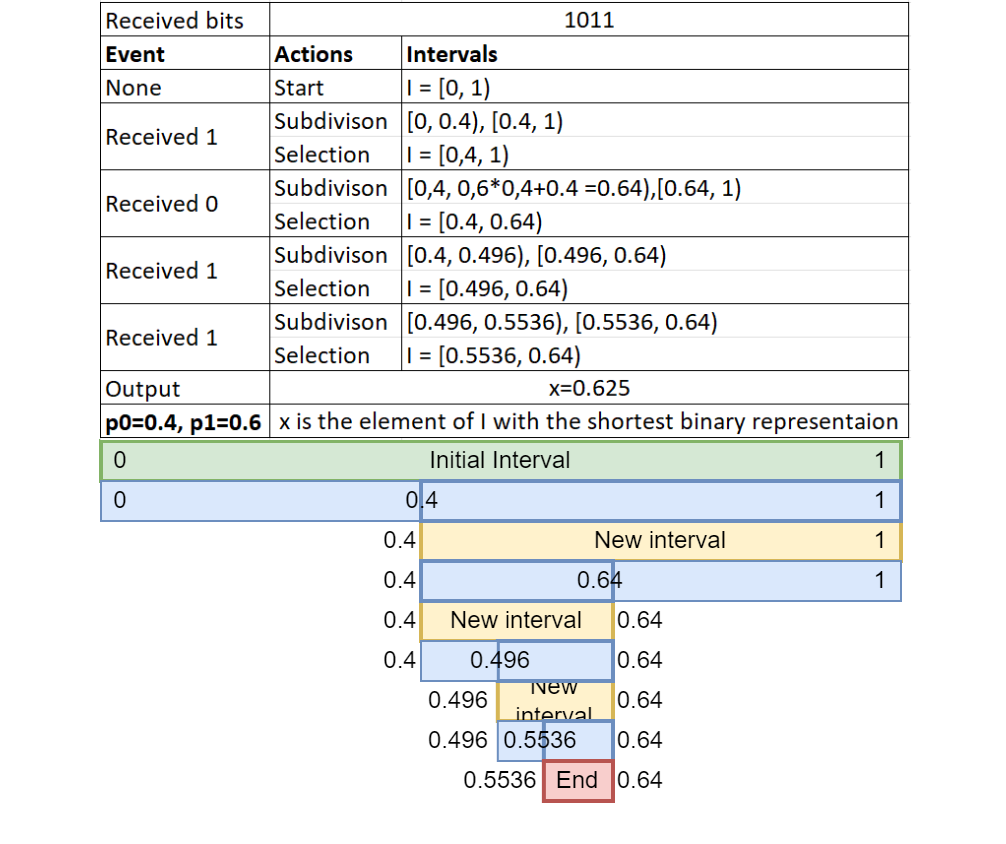}
    \vspace{-5mm}
    \caption{\footnotesize{Example of an arithmetically coded bitstream with a fixed probability of appearance of 0 and 1}}
    \label{arithmetic-coding-example}
\vspace{-2mm}
\end{figure}

An example of a simple bitstream coded with an arithmetic coder is depicted in Fig. \ref{arithmetic-coding-example}.

\vspace{-2mm}
\section{Arithmetic coding adapted to constrained DNA codes}
\vspace{-2mm}
To adapt the coder to DNA data storage, like in the JPEG 2000 algorithm, the coefficients are represented into bit planes and then arithmetically encoded with successively included intervals that are computed according to the statistics of the input bits. Finally, the interval representing the whole bit-plane representation is encoded into a DNA constrained code instead of a binary code.
In order for the coder to be working, a rule has to be respected: To be able to encode any interval output by the arithmetic coder, the set of numbers that can be coded with constrained DNA codes must be dense in the interval $[0, 1]$. 

\vspace{-4mm}
\subsection{Constrained code}
\vspace{-2mm}


As previously mentioned, respecting the biochemical constraints identified by the state of the art, helps preventing the apparition of errors along the process. In this solutions we will only focus on one type of biochemical constraint: prevent the apparition of homopolymers in the codes (meaning that a nucleotide will not be repeated too many times in a row, generally 3 repetitions is the maximum accepted). 
\begin{figure}
    \centering
    \begin{tikzpicture} [node distance = 2cm, on grid]
        \node (q0) [state, initial, accepting] {$\Sigma_4$};
        \node (q1) [state, accepting] at (2,0) {$\Sigma_4$};
        \node (q2) [state, accepting] at (6,0) {$\Sigma_4\backslash\lbrace{x}\rbrace$};
        \path [-stealth, thick]
            (q0) edge [bend left] node [above] {$x$} (q1)
            (q1) edge [bend left] node[above] {$x$} (q2)
            (q2) edge [bend left] node[below] {$x$}  (q0);
    \end{tikzpicture}
    \vspace{-4mm}
    \caption{\footnotesize{Automata defining the coding alphabets used to encode the interval into DNA}}
    \label{fig:my_label}
\vspace{-2mm}
\end{figure}
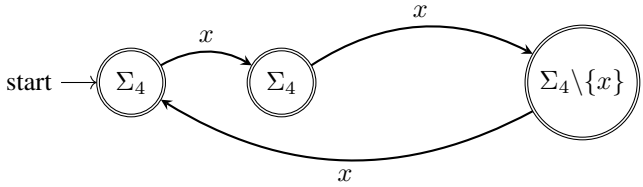

This generates a code $C_3$ with codewords of length 3 that can be associated to symbols in a dictionary defined as follows:
\\ 
$C_3 = \lbrace AAT, AAC, AAG, ATA, ATC, ATG, ACA, \hspace{0.1cm}..., \\\hspace{0.1cm} GCT, GCG, GGA, GGT, GGC\rbrace$,
\\
$|C_3|=48$,
\\
$D = \lbrace(k,c)\slash (k\in\llbracket0,47\rrbracket \land c\in C_3)\rbrace \land$\\$ \forall(k_1,c_1)\in D, \forall(k_2,c_2)\in D, (k_1=k_2\Rightarrow c_1=c_2)$.\\

This dictionary will be the base for our interval coder, that will consist in writing numbers in base 48 and translating the numeric development with the dictionary.
For that, it is needed to prove that this system can code a set of numbers that is dense in $[0,1]$.
\begin{proposition}
The set of encodable elements using the dictionary $D$ is dense in $[0, 1]$.
\end{proposition}
\begin{proof}
Let $S$ be the following set: $S = \{ \frac{a}{48^n}, (a, n)\in \mathbb{N}^2 \}$\\
Let $x \in [0, 1]$ \\
Let $(x_n)_{n\in\mathbb{N}}\in S^\mathbb{N}$, with
$\forall n \in \mathbb{N}, x_n = \frac{\lfloor 48^nx\rfloor}{48^n}$\\
$\forall n \in \mathbb{N}, 48^nx\leq \lfloor 48^nx\rfloor < 48^nx + 1$\\
$\Rightarrow \forall n \in \mathbb{N}, x\leq \frac{\lfloor 48^nx\rfloor}{48^nx} < x + \frac{1}{48^n}$\\
$\Rightarrow \forall n \in \mathbb{N}, x\leq x_n < x + \frac{1}{48^n}$\\
$\Rightarrow \lim_{x\to\infty} x_n = x$\\
$\Rightarrow \forall x \in [0, 1], \exists (x_n)_{n\in\mathbb{N}}\in S^\mathbb{N} / \lim_{x\to\infty} x_n = x$\\
So S is dense in $[0, 1]$
\end{proof}

To encode an interval, the number inside the interval with the shortest development in base 48 is picked.

\vspace{-2mm}
\section{Integration into JPEG 2000}
\vspace{-2mm}
\subsection{Principle}
\vspace{-2mm}

The modified JPEG 2000 image codec encodes bit plane representation of the DWT coefficients into a quaternary stream. The adaptation of the JPEG 2000 algorithm is done as follows: an image is first transformed into coefficients that are quantized and bit-plane represented. It non longer uses the classic MQ-coder to encode the bit-plane representations of the DWT coefficients, but the proposed DNA arithmetic coder. The rest of the DNA encoding consists in transcoding byte per byte the binary overheads present in the standard JPEG 2000 files.
\setlist{nolistsep}
\begin{itemize}[noitemsep]
    \item The overheads are directly transcoded byte per byte using a fixed length quaternary coder
    \item The data usually encoded by the MQ-coder is encoded using our new DNA-like data adapted arithmetic coder.
\end{itemize}
Both of these inputs are merged into a single quaternary stream. This stream is then cut into oligos of length 200, which is a biochemical limits for error-controlled DNA synthesis. Oligo specific overheads are developed to ensure decodability of the data. An index is encoded in each oligo to represent its position in the general oligo pool so that at decoding, the general stream can be reconstructed properly. 
\vspace{-3mm}
\subsection{Development}
\vspace{-2mm}
The OpenJPEG software\footnote{\color{blue}\url{http://www.openjpeg.org/}} was used as a baseline for the development of our software. A private C library called openjpdna was developed to make the link between the core of our solution (the arithmetic coder) and the modified software. The modification of the original software was operated by consulting and writing into the place of the MQ-coder. Because the overheads are transcoded into DNA without being interpreted, the proposed design is an open-loop modification of the JPEG 2000 codec. 

%% file: results.tex
\vspace{-2mm}
\section{Experimental results}
\begin{figure}[ht]
\centering
\vspace{-1\baselineskip}
	\begin{minipage}[b]{0.45\linewidth}
		\centerline{
            \includegraphics[width = 1.2in]{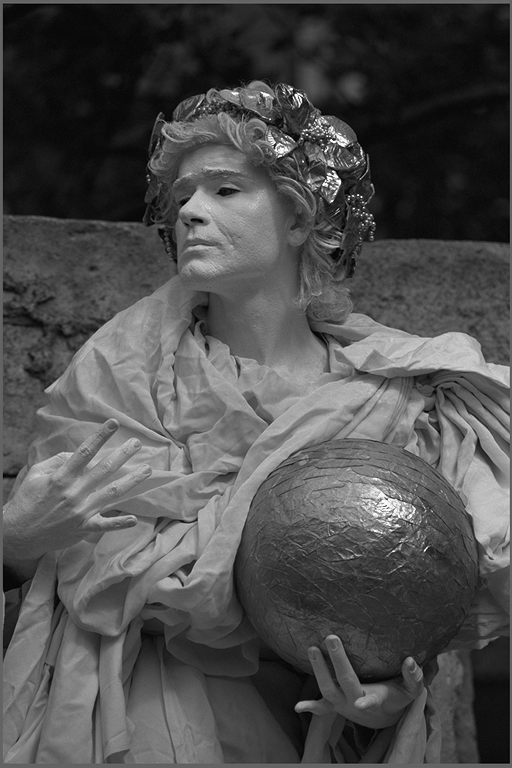}}
            \centering{\footnotesize{(a) Input image}}
	   \end{minipage}
	\begin{minipage}[b]{0.45\linewidth}
		\centerline{
            \includegraphics[width = 1.2in]{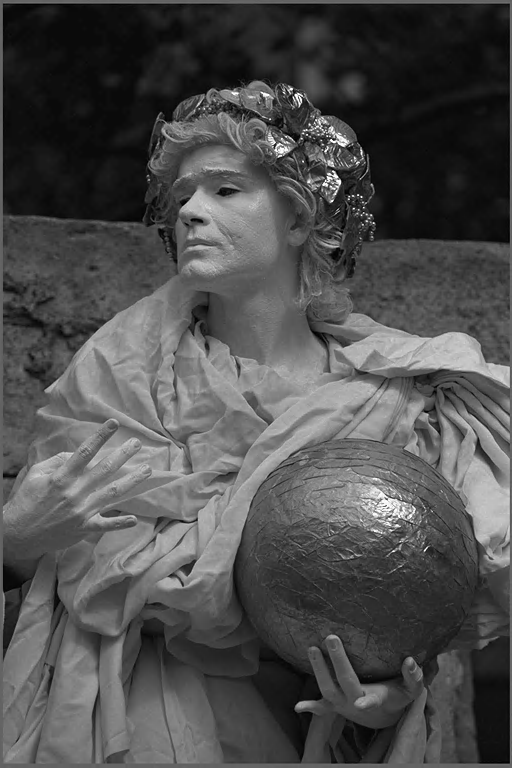}}
            \centering{\footnotesize{(b) Decoded image at 35dB}}
	   \end{minipage}
\vspace{-0.5\baselineskip}
\caption{\footnotesize{Visual results on the compression of kodim17}}
\vspace{-1\baselineskip}
\label{fig:kodak}
\end{figure}

\begin{figure}[h]
    \centering
    \includegraphics[width = 3.5in]{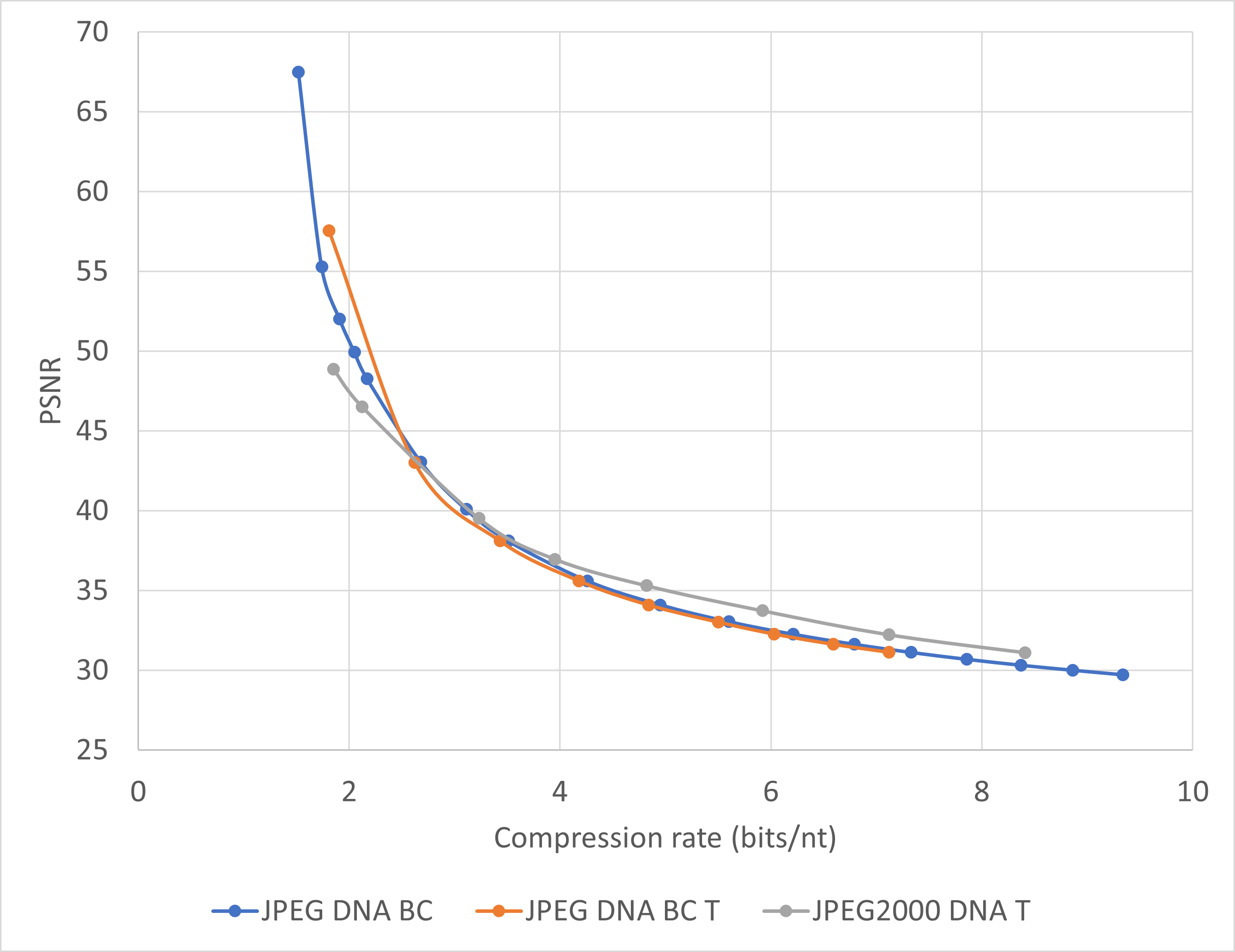}
\vspace{-1.5\baselineskip}
    \caption{\footnotesize{Comparison of the rate distorsion for kodim17 on JPEG2000DNA, JPEG DNA BC and JPEG DNA BC Transcoder}}
    \label{fig:perftests}
\vspace{-1\baselineskip}
\end{figure}

\vspace{-3mm}
The proposed software's performance was evaluated on 8bpp gray-level versions of the kodak dataset\footnote{\color{blue}\url{http://r0k.us/graphics/kodak/}}. The results were then compared to two other image codecs for synthetic DNA, JPEG DNA BC \cite{Dimopoulou}, JPEG DNA BC Transcoder \cite{JPEGDNABCT}, based on \cite{Dimopoulou}. Visual examples of the obtained results can be seen in Fig.\ref{fig:kodak} and performance results in Fig.\ref{fig:perftests}. As the performance results show in Fig.\ref{fig:perftests}, the novel methods has improved performance over the previously used methods (JPEG DNA BC and JPEG DNA BC Transcoder).


%% file: conclusion.tex
\vspace{-0.7cm}
\section{Conclusion}
\vspace{-3.5mm}
In this paper, we propose a novel quaternary constrained arithmetic coder adapted to synthetic DNA data storage. This coder respects biochemical constraints defined by biochemistry to improve the compression and reduce the number of errors of the biochemical operations. We integrated this coder into a binary image codec that used a binary arithmetic coder, JPEG2000. The resulting quaternary image codec shows improved performance over the previous image codecs for synthetic DNA data storage. A more thorough implementation is necessary to improve the compression performance by integrating the arithmetic coder inside the JPEG 2000 in a closed loop manner. Improving the coding of the overheads might also improve the performance of the image codec. Studying the influence of DNA data storage errors will also give insight on the direction to take to further improve the robustness of the codec.

%% file: main.bbl
\begin{thebibliography}{10}

\bibitem{Church}
George~M. Church, Yuan Gao, and Sriram Kosuri,
\newblock ``Next-generation digital information storage in dna,''
\newblock {\em Science}, vol. 337, no. 6102, pp. 1628--1628, 2012.

\bibitem{Goldman2013}
N.~Goldman, P.~Bertone, and S.~Chen,
\newblock ``Towards practical, high-capacity, low-maintenance information
  storage in synthesized dna,''
\newblock {\em Nature}, 2013.

\bibitem{Aeon}
Marius Welzel, Peter~Michael Schwarz, Hannah~F. Löchel, Tolganay Kabdullayeva,
  Sandra Clemens, Anke Becker, Bernd Freisleben, and Dominik Heider,
\newblock ``Dna-aeon provides flexible arithmetic coding for constraint
  adherence and error correction in dna storage,''
\newblock {\em Nature Communications}, 2023.

\bibitem{JPEG2000}
David~S. Taubman and Michael~W. Marcellin,
\newblock ``Jpeg2000 image compression fundamentals, standards and practice,''
\newblock {\em Springer}.

\bibitem{DNARAM}
S.~M. Hossein~Tabatabaei Yazdi, Ryan Gabrys, and Olgica Milenkovic,
\newblock ``Portable and error-free dna-based data storage,''
\newblock {\em Nature}, 2017.

\bibitem{Welzel}
Michael Schwarz, Marius Welzel, Tolganay Kabdullayeva, Anke Becker, Bernd
  Freisleben, and Dominik Heider,
\newblock ``Mesa: automated assessment of synthetic dna fragments and
  simulation of dna synthesis, storage, sequencing and pcr errors,''
\newblock {\em Bioinformatics}, vol. 36, pp. 3322--3326, 2020.

\bibitem{Imperial}
Jamie~J. Alnasir, Thomas Heinis, and Louis Carteron,
\newblock ``Dna storage error simulator: A tool for simulating errors in
  synthesis, storage, pcr and sequencing,''
\newblock .

\bibitem{DNASmart}
Chisom Ezekannagha, Marius Welzel, Dominik Heider, and Georges Hattab,
\newblock ``Dnasmart: Multiple attribute ranking tool for dna data storage
  systems,''
\newblock {\em Computational and Structural Biotechnology Journal}, vol. 21,
  2023.

\bibitem{DNAcoding}
Melpomeni Dimopoulou, Marc Antonini, Pascal Barbry, and Raja Appuswamy,
\newblock ``A biologically constrained encoding solution for long-term storage
  of images onto synthetic dna,''
\newblock {\em European Signal Processing Conference (EUSIPCO)}, 2019.

\bibitem{JPEGDNABCT}
Luka Secilmis, Michela Testolina, Davi Nachtigall~Lazzarotto, and Touradj
  Ebrahimi,
\newblock ``Towards effective visual information storage on dna support,''
\newblock {\em Applications of Digital Image Processinf XLV}, 2022.

\bibitem{SFC4}
Xavier Pic and Marc Antonini,
\newblock ``A constrained shannon-fano entropy coder for image storage in
  synthetic dna,''
\newblock {\em European Signal Processing Conference (EUSIPCO)}, 2022.

\bibitem{MWCC2022Iulia}
Iulia Mitrica, Xavier Pic, Eva~Gil San~Antonio, Melpomeni Dimopoulou, and Marc
  Antonini,
\newblock ``Efficient classification of dna reads for robust decoding of data
  stored in synthetic dna,''
\newblock {\em Munich Workshop on Coding and Cryptography (MWCC 2022)}, 2022.

\bibitem{MWCC2022Xavier}
Xavier Pic and Marc Antonini,
\newblock ``Image coding algorithm for dna data storage combining jpeg and
  autoencoders,''
\newblock {\em Munich Workshop on Coding and Cryptography (MWCC 2022)}, 2022.

\bibitem{DWT}
Marc Antonini, Michel Barlaud, Pierre Mathieu, and Ingrid Daubechies,
\newblock ``Image coding using wavelet transform,''
\newblock {\em IEEE Transactions on Image Processing}, vol. 1, pp. 205--220,
  1992.

\bibitem{EBCOT}
Chengyi Xiong, Jianhua Hou, Zhirong Gao, and Xiang He,
\newblock ``High performance scalable image compression with ebcot,''
\newblock {\em IEEE International Conference on Multimedia and Expo}, 2007.

\bibitem{Dimopoulou}
Melpomeni Dimopoulou, Eva~Gil San~Antonio, and Marc Antonini,
\newblock ``A jpeg-based image coding solution for data storage on dna,''
\newblock {\em European Signal Processing Conference (EUSIPCO)}, 2021.

\end{thebibliography}
